# Waveguide–based Electroabsorption Modulator Performance

**RUBAB AMIN[1], JACOB B. KHURGIN[2], VOLKER J. SORGER[1,*]**

[1]*Department of Electrical and Computer Engineering, George Washington University*
*800 22nd St., Science & Engineering Hall, Washington, DC 20052, USA*
[2]*Department of Electrical and Computer Engineering, Johns Hopkins University,*
*Baltimore, Maryland 21218, USA*
*[*]sorger@gwu.edu*

**Abstract:** Electro-optic modulation is a key function for data communication. Given the vast amount of data handled, understanding the intricate physics and trade-offs of modulators on-chip allows revealing performance regimes not explored yet. Here we show a holistic performance analysis for waveguide-based electro-absorption modulators. Our approach centers around material properties revealing obtainable optical absorption leading to effective modal cross-section, and material broadening effects. Taken together both describe the modulator physical behavior entirely. We consider a plurality of material modulation classes to include two-level absorbers such as quantum dots, free carrier accumulation or depletion such as ITO or Silicon, two-dimensional electron gas in semiconductors such as quantum wells, Pauli blocking in Graphene, and excitons in two-dimensional atomic layered materials such as found in transition metal dichalcogendies. Our results show that reducing the modal area generally improves modulator performance defined by the amount of induced electrical charge, and hence the energy-per-bit function, required switching the signal. We find that broadening increases the amount of switching charge needed. While some material classes allow for reduced broadening such as quantum dots and 2-dimensional materials due to their reduced Coulomb screening leading to increased oscillator strengths, the sharpness of broadening is overshadowed by thermal effects independent of the material class. Further we find that plasmonics allows the switching charge and energy-per-bit function to be reduced by about one order of magnitude compared to bulk photonics. This analysis is aimed as a guide for the community to predict anticipated modulator performance based on both existing and emerging materials.

## 1. Introduction

Applications in data centers (core computing) and the looming era of sensor-driven internet-of-things (edge computing) drive demand for data bandwidth. Given the limitation of charging electrical wires, the need for optical communication is made [1]. Thus, electrical control of optical signals (electro-optic, EO) conversion remains a critical function [2]. However, the weak light–matter interaction (LMI) of Silicon or III-V photonics electro-optic modulators (EOM) requires footprints, energy-per-bit functions that are orders of magnitude higher compared to their electronic 3-termical switching counterparts (i.e. transistors). The challenge is to obtain the highest optical effective mode index change with the least amount of voltage (i.e. steepest switching), while observing optical loss limitations [3,4], whereas the latter is a fundamental physics constrain given by the Kramers-Kronig relationship. While EO modulation can be enabled by either changing the real part ($n$) of the modal refractive index leading to phase shifting-based interferometer-like devices termed here electro-optic modulators (EOM), this work focuses solely on physical effects modulating the imaginary part ($\kappa$) leading to the class of electro-absorptive modulators (EAM). In both types, the fundamental complex index of refraction is altered electrically inside the active material, modifying the optical mode's propagation constant inside the waveguide. Assuming on-off-keying (OOK), EOMs always require an interferometric scheme to induce an amplitude modulation which fundamentally requires more real estate on-chip compared to EAMs, which can induce optical amplitude changes directly in a linear device design. Here we make the distinction between current-driven and voltage-driven modulators; in the former the index change is a result of injecting carriers into (or removing from) the active region thus enabling (disabling) the optical transitions and therefore adding (subtracting) to the oscillators strength inside the active material layer. We only consider current-driven (i.e. 'gated') modulation effects in this work. In contrast, in voltage-driven modulators no current flows in/out of the active region, and the change in index is evoked by the energy level shifts and oscillator strength change induced by the electric field (e.g. Stark effect). The widely used lithium niobate (LiNbO$_3$) modulators based on Pockels effect and III-V semiconductor based quantum-confined Stark effect modulators are two examples of voltage driven EO modulators [5-7]. The spectrum of active materials and their respective modulation effects is wide and includes III-V based a) QDs and b) QWs, free carrier based c) Si and d) ITO, and emerging materials such as atomically-thin e) Graphene and f) transition metal dichalcogenides (TMD). While promising device performance has been made in an ad-hoc manner, a systematic analysis and subsequent performance comparison between optical mode, active material mechanism, and subsequent device performance tradeoffs is yet outstanding. Here we show such comparison for the first time where we contrast the different modulation mechanisms-material combinations, to include III-V based band-filling a) QDs and b) QWs, free carrier based c) Si and d) ITO, and novel 2D materials such as e) Graphene's Pauli Blocking and f) exciton modulation in transition metal dichalcogenides (TMD). We show that the modulation performance based on free-carriers falls show of all other mechanisms considered, while remaining materials perform about equal. For the latter, we further show that the energy-bandwidth-ratio (EBR), an EAM figure of merit, strongly depends on both the effective mode index and broadening effects.

## 2. Absorption Modulator Definitions

### 2.1 Effective area of the waveguide

Key to the modulator's performance is a strong LMI which requires a small effective mode area, $S_{eff}$, or simply in one dimension (1D), a short effective length, $t_{eff}$. To determine the effective thickness of the waveguide we evaluate the Poynting vector, $S(x) = E_y H_x - E_x H_y$. According to the Maxwell equations $\frac{\partial}{\partial z} H_x = \beta H_x = \omega n^2(x) \varepsilon_0 \varepsilon_{eff} E_x$ and $\frac{\partial}{\partial z} H_y = \beta H_x = -\omega n^2(x) \varepsilon_0 \varepsilon_{eff} E_x$. Thus, $S(x) = \frac{\omega n^2(x,y) \varepsilon_0 \varepsilon_{eff} (E_y^2 + E_x^2)}{2\beta} = \frac{n^2(x,y) \varepsilon_{eff}(E_y^2 + E_x^2)}{2 n_{eff} \eta_0}$, where the effective index has been introduced as $n_{eff} = \beta c / \omega$. The total power flow is then

$$P = \int_{-\infty}^{\infty}\int_{-\infty}^{\infty} S(x,y) dxdy = \frac{1}{2 n_{eff} \eta_0} \int_{-\infty}^{\infty}\int_{-\infty}^{\infty} n^2(x,y)\left(E_x^2 + E_y^2\right) dxdy = \frac{n_{eff} E_{a0}^2}{2\eta_0} S_{eff}$$ where $E_{a0}$ is the

magnitude of the transverse electric field in the middle of active layer (Fig.1). The effective area of the waveguide is

$$S_{eff} = \frac{1}{n_{eff}^2} \int_{-\infty}^{\infty}\int_{-\infty}^{\infty} n^2(x,y)\left(E_x^2 + E_y^2\right) dxdy / E_{a0}^2 \qquad (1)$$

This definition of the effective area may differ from others in the literature, but the difference is small and involves the distinction between the effective and group indices. We note that this definition includes the fact that active layer is not necessary at the center of the waveguide, i.e. $E_a$ may not be the peak electric field.

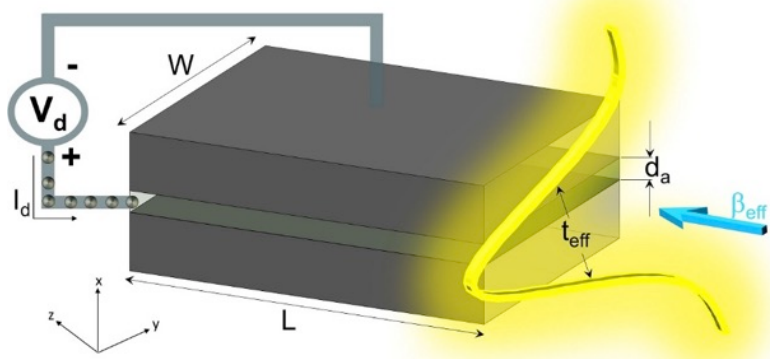

Fig. 1: Schematic of the waveguide structure with a biasing scheme to the active region (drive voltage, $V_d$ and current (charge) flow into the active layer, $I_d$) and propagation direction (indicated with blue arrow, $\beta_{eff}$). The associated electric field in the y-direction is shown. $d_a$ is the width of the active region and $t_{eff}$ is the effective thickness. The relevant coordinate system used in this work is also included to accompany the text.

## 2.2 Absorption cross-section and effective thickness

Let us now estimate the absorption change induced by the injection (or depletion of carriers). We first consider two level absorbers, which can be atoms or quantum dots (QD's). The absorbers have Lorentzian line shape with linewidth (broadening) $\gamma$ and according to Fermi's Golden rule the density of photons changes as

$$\frac{dN_{ph}(x,y)}{dt} = \frac{2\pi}{h}\frac{e^2}{4}E_{eff}^2 r_{12}^2 \rho(E) N_a(x,y) \tag{2}$$

where we have defined $E_{eff}$ as the component of the electric field that is actually being absorbed (for many 2D structures it is the in-plane or TE component) and where $r_{12}$ is the transition matrix element and the factor of ¼ comes from the fact that only positive frequencies cause the transition. The density of states is

$$\rho(E) = \frac{1}{\pi}\frac{h\gamma}{h^2\Delta\omega^2 + h^2\gamma^2} \tag{3}$$

where $\Delta\omega = \omega - \omega_0$ is the detuning from the resonance. Then the change of the energy density due to absorption in the active layer is

$$\frac{dU(x,y)}{dt} = h\omega\frac{dN_{ph}(x,y)}{dt} = e^2 r_{12}^2 \frac{E_{eff}^2(x,y)}{2}\frac{\omega}{h\gamma}\frac{\gamma^2}{\Delta\omega^2+\gamma^2} N_a(x,y) \tag{4}$$

Integrating over the cross-section we obtain

$$\frac{dP}{dz} = -\int_{active}\int_{-W/2}^{W/2}\frac{dU}{dt}dx = \frac{-e^2 r_{12}^2}{2}\frac{\omega}{h\gamma}\frac{\gamma^2}{\Delta\omega^2+\gamma^2}\int_{active}\int_{-W/2}^{W/2} N_a(x,y)E_{eff}^2(x,y)dxdy \tag{1}$$

Now we can introduce the effective area of the active region as

$$S_{a,eff} = \int_{active}\int_{-W/2}^{W/2} N_a E_{eff}^2 dxdy / E_{a0}^2 N_{a0} \tag{2}$$

and re-write (1) using **Error! Reference source not found.** and (2)

$$\frac{dP}{dz} = -\frac{e^2 r_{12}^2 \eta_0}{n_{eff}}\frac{\omega}{h\gamma}\frac{\gamma^2}{\Delta\omega^2+\gamma^2}\frac{S_{a,eff}}{S_{eff}}N_{a0}P = -\sigma_a(\omega)\frac{S_{a,eff}}{S_{eff}}N_{a0}P \tag{3}$$

where the absorption cross-section has been introduced as

$$\sigma_a(\omega) = \frac{e^2 \eta_0}{h n_{eff}} r_{12}^2 \frac{\omega}{\gamma} \frac{\gamma^2}{\Delta\omega^2 + \gamma^2} = \frac{4\pi\alpha_0}{n_{eff}} r_{12}^2 \frac{\omega}{\gamma} L(\omega) \qquad (4)$$

Here $\alpha_0$ is the fine structure constant and $L(\omega) = \gamma^2 / (\Delta\omega^2 + \gamma^2)$ represents a Lorentzian line shape that has maximum value at resonance equal to unity. The absorption coefficient is therefore

$$\alpha(\omega) = \sigma_a(\omega) \frac{S_{a,eff}}{S_{eff}} N_{a0} \qquad (5)$$

The significance of this expression is that the absorption is clearly separated into two factors – intrinsic (or material) the absorption cross-section) and the geometrical (or confinement) factor, $\Gamma = S_{a,eff} / S_{eff}$.

Now, often the active layer is essentially two dimensional, with two-dimensional density of carriers that is independent of the lateral direction, i.e. $N_{2D} = \int_{active} N_a(x) dx$. Then we can find effective active area (2) as

$$S_{a,eff} = \frac{N_{2D}}{N_{a0}} W F \qquad (6)$$

Where the "uniformity function" is

$$F = \int_{active} \frac{N_a(x)}{N_{a0}} \int_{-W/2}^{W/2} \frac{E_{eff}^2(x,y)}{E_{a0}^2} dy dx \bigg/ W \int_{active} \frac{N_a(x)}{N_{a0}} dx \qquad (7)$$

Clearly, when the active layer is essentially a delta function, $N_a(x) = N_{2D} \delta(x - x_a)$

$$F = W^{-1} \int_{-W/2}^{W/2} \frac{E_{eff}^2(x_a, y)}{E_{a0}^2} dy \qquad (8)$$

and if the field is uniform in the lateral direction $F = E_{eff}^2 / E_a^2 = \cos^2 \theta_{eff}$, where $\theta_{eff}$ is the angle between the dipole and electric field. Now we can write (5) as

$$\alpha(\omega) = \sigma_a(\omega) \frac{WF}{S_{eff}} N_{2D} = \frac{\sigma_a(\omega)}{t_{eff}} N_{2D} \qquad (9)$$

where the effective thickness of the waveguide is

$$t_{eff} = S_{eff} / WF = S'_{eff} / W \qquad (10)$$

where $S'_{eff} = S_{eff} / F$ For the waveguides in which the field does not change much laterally which is often the case,

$$t_{eff} \approx \frac{1}{n_{eff}^2} \int_{-\infty}^{\infty} n^2(x,y)\left(E_x^2 + E_y^2\right) dx / E_{a0}^2 \qquad (11)$$

*2.3 Switching characteristics: Definitions*

Next, we introduce the absorption modulation characteristics. Here we (arbitrarily) select the maximum (minimum) transmission to be 90% (10%), for an extinction ratio (ER) of ~9.5dB modulation. Minimum transmission (maximum absorption) is achieved when no electrons are injected

$$\alpha_{max}(\omega)L = \sigma(\omega)N_{2D}L/t_{eff} = \ln(10) \approx 2.302 \tag{12}$$

whereas maximum transmission occurs when $\delta n_{2D}$ carriers are injected by applying the positive voltage to the gate,

$$\alpha_{min}(\omega)L = \sigma(\omega)(N_{2D} - \delta n_{2D})L/t_{eff} = -\ln(0.9) \approx 0.105 \tag{13}$$

Subtracting (14) from (13) we obtain

$$\delta n_{2D}L \approx 2.2 t_{eff}/\sigma(\omega) \tag{14}$$

and when multiplied by the waveguide width and length, we obtain the expression and using (11) we can evaluate the key modulator characteristics – the switching charge

$$Q_{sw} = eWL\delta n_{2D} = 2.2eWt_{eff}/\sigma(\omega) = 2.2eS_{eff}/\sigma(\omega)F \tag{15}$$

So, in the end the switching charge depends only on the ratio of the effective cross-section of the waveguide and the "effective absorption cross-section" of the material $\sigma(\omega)F$. One can then determine the switching voltage as

$$V_{sw} = \frac{d_{gate}e\delta n_{2D}}{\varepsilon_0\varepsilon_{eff}} = 2.2\frac{e}{\varepsilon_0\varepsilon_{eff}}\frac{t_{eff}}{L}\frac{d_{gate}}{\sigma(\omega)} \approx \frac{e}{\varepsilon_0\varepsilon_{eff}}N_{2D}d_{gate} \tag{16}$$

Where $d_{gate}$ and $\varepsilon_{eff}$ are the thickness and dielectric constant of the insulator between the active layer and the gate. Note that voltage depends only on the density of the QDs and thus can be adjusted within reasonable limits. Similarly, according to (12) we have the freedom of choosing the length of the modulator as long as

$$L \approx 2.302 t_{eff}/N_{2D}\sigma(\omega) \tag{17}$$

Therefore, the capacitance of the modulator can be found as

$$C = \varepsilon_0\varepsilon_{eff}WL/d_{gate} = 2.3\frac{\varepsilon_0\varepsilon_{eff}}{d_{gate}}\frac{Wt_{eff}}{N_{2D}\sigma_a} = 2.3\frac{\varepsilon_0\varepsilon_{eff}}{d_{gate}}\frac{S_{eff}}{N_{2D}\sigma_a F} \tag{18}$$

The 3dB cut-off frequency of the modulator then becomes

$$f_{3dB} = 1/2\pi RC \approx \frac{d_{gate}}{\varepsilon_0\varepsilon_{eff}}\frac{N_{2D}\sigma_a F}{7.2 S_{eff} R} \tag{19}$$

Finally, we can determine the switching energy (per bit) as

$$U_{sw} = \frac{1}{2}Q_{sw}V_{sw} \approx \frac{e^2 d_{gate}}{\varepsilon_0\varepsilon_{eff}}\frac{N_{2D}S_{eff}}{\sigma_a(\omega)F} \tag{20}$$

Also we can define the relevant figure of merit – the ratio of switching energy to the 3dB cut-off frequency or energy-bandwidth-ratio as

$$EBR = U_{SW} / f_{3dB} \approx 7.2 e^2 R \frac{S_{eff}^2}{\sigma_a^2(\omega) F^2} \tag{21}$$

This expression is strikingly simple as it does not depend on anything but the intrinsic absorption strength of the active material, $\sigma_a(\omega)$, degree of waveguide confinement, $S_{eff}$ and the efficiency of coupling between the electric field in the waveguide and the active medium, $F$. First we shall investigate the intrinsic absorption strengths by considering five different media used or proposed for absorption modulation.

## 3. Material Classes for EAMs

### 3.1 Modulator based on two-level absorbers

The first class of modulators is the one based on two-level absorbers, which for electrically driven modulator indicates semiconductor driven QDs. The absorption cross-section in QDs has been already determined in (4) and now we can evaluate it using the relation between the dipole matrix element, $r_{12}$ and the matrix element of the momentum for the interband transition, $P_{cv}$ as

$$r_{12} = \sqrt{\frac{F_{cv}}{2}} \frac{P_{cv}}{m_0 \omega} \tag{22}$$

where the term in front includes the overlap between the "envelope" wave-functions of the conduction and valence bands $F_{cv} = \left| \int \psi_c \psi_v dV \right|^2 <\sim 1$ and the factor of ½ associated with specifics of the heavy hole wave-function involved in the transition. Furthermore, one can use the relation between the above momentum matrix element and the electron effective mass, $m_c$ as

$$\frac{m_0}{m_c} = 1 + \frac{2 P_{cv}^2}{m_0 E_g} \approx \frac{E_p}{E_g} \tag{23}$$

where for wide range of direct bandgap semiconductors $E_P = 2 P_{cv}^2 / m_0 \approx 16 - 24 eV$ and $E_g \approx \hbar\omega$ is the bandgap energy. It is not difficult then to obtain the expression for the absorption cross-section on resonance

$$\sigma_{QD}(\omega) = \frac{\pi\alpha_0}{n_{eff}} F_{cv} \frac{\hbar}{\gamma}\left(m_c^{-1} - m_0^{-1}\right) = \frac{\pi\alpha_0}{n_{eff}} F_{cv} \frac{\hbar^2 / m_0}{\hbar\gamma}\left(\frac{m_0}{m_c} - 1\right) = 1.75 \times 10^{-17} cm^2 \frac{F_{cv}}{\hbar\gamma n_{eff}}\left(\frac{m_0}{m_c} - 1\right) \tag{24}$$

where the energy broadening $\hbar\gamma$ is given in electron volts (eV). As one can perceive, the absorption cross-section does

$$\sigma_{QD}(\omega) = \frac{\pi\alpha_0}{n_{eff}} F_{cv} \frac{E_P}{\hbar\omega} \frac{\hbar^2 / m_0}{\hbar\gamma} \approx 3.5 \times 10^{-16} cm^2 \frac{F_{cv}}{(\hbar\gamma)(\hbar\omega) n_{eff}} \tag{25}$$

where all the energies are in eV. One can further show that if we introduce additional inhomogeneous broadening, possibly important in QDs, but less so for doped ions, it will simply add up as roughly $\gamma \approx \sqrt{\gamma_{Lorentz}^2 + \gamma_{inh.}^2}$ (Voigt profile), and one can still use the expressions (24) and (25). Thus in the end, absorption cross-section depends significantly only on the broadening. Finally, using (17) and (24) we obtain the required length of QD modulator

$$L_{QD} \approx 2.3 \frac{n_{eff} t_{eff}}{\pi \alpha_0 N_{QD}} \frac{\gamma}{F_{cv} \mathrm{h}\left(m_c^{-1} - m_0^{-1}\right)} \tag{26}$$

*3.2 Free carrier accumulation- depletion style modulator*

Let us consider now the free electron gas in the conduction band of a semiconductor which can be Si, ITO, or some other material whose density is $N_e(x, y)$. Dielectric constant in Drude approximation can be written as

$$\varepsilon_r(\omega, x, y) = \varepsilon_\infty - \frac{N_e(x, y)e^2}{\varepsilon_0 m_c} \frac{1}{\omega^2 + i\omega\gamma} \tag{27}$$

where $\varepsilon_\infty$ is the dielectric constant of the undoped semiconductor. The rate of absorption of electro-magnetic energy is proportional to the imaginary part of the dielectric constant as

$$\frac{dU(x,y)}{dt} = \omega \varepsilon_0 \varepsilon_{im}(\omega, x) \frac{E_{eff}^2(x)}{2} = \frac{N_e(x,y)e^2}{m_c} \frac{\gamma}{\omega^2 + \gamma^2} \frac{E^2(x,y)}{2} \tag{28}$$

where effective electric field, $E_{eff}$ is the same as the total field, $E$. This expression is similar to **Error! Reference source not found.**, so following the familiar steps will lead us to (5) with the effective absorption area described by (2) with $N_e$ in place of $N_a$ and absorption cross-section now defined as

$$\sigma_{fc}(\omega) = \frac{4\pi\alpha_0}{n_{eff}} \frac{\mathrm{h}}{m_c} \frac{\gamma}{\omega^2 + \gamma^2} = \frac{4\pi\alpha_0}{n_{eff}} \frac{\mathrm{h}^2}{m_0(\mathrm{h}\gamma_{eff})} \frac{m_0}{m_c} \approx \frac{7.02 \times 10^{-17} cm^2}{(\mathrm{h}\gamma_{eff})} \times \frac{m_0}{m_c} \tag{29}$$

where the effective detuning is $\gamma_{eff} = \gamma + \omega^2/\gamma$. As one can see this expression is similar to the one for the QDs with one major difference – due to non-resonant character of the absorption the cross section for the free carriers is orders of magnitude lower than for the resonant QDs. Also, it is easy to see that for the wavelength in the telecom range the cross-section actually increases with the increase of broadening caused by scattering $\gamma$. Therefore, as shown in the next section, ITO is intrinsically a better material than high quality Si. Note that, the maximum absorption is achieved at $\gamma = \omega$ when $\gamma_{eff} = 2\omega$ and

$$\sigma_{fc.\max}(\omega) = \frac{4\pi\alpha_0}{n_{eff}} \frac{\mathrm{h}}{m_c \omega} \times \frac{1}{2} = 3.53 \times 10^{-17} cm^2 \frac{1}{\mathrm{h}\omega n_{eff}} \frac{m_0}{m_c} \tag{30}$$

*3.3 Two-dimensional electron gas in semiconductor QWs*

Let us consider an $N_{QW}$ quantum well inside the waveguide, where the each QW is populated with the 2D carrier density $n_{2D}$ that are distributed according to Fermi Dirac distribution

$f_c(E_c) = \{1 + \exp[(E_c - E_f)/kT]\}^{-1}$ and the Fermi energy (relative to the bottom of conduction band) can be found as

$$E_f = kT \ln\left[\exp\left(\frac{\pi \hbar^2 n_{2D}}{N_{QW} m_c kT}\right) - 1\right] \tag{31}$$

which can be differentiated to obtain

$$\frac{dn_{2D}}{dE_f} = \frac{N_{QW} m_c}{\pi \hbar^2} \frac{1}{1 + \exp(-E_f/kT)} \tag{32}$$

Let us now evaluate the absorption rate of the carriers; for that following Fermi's Golden rule we introduce the rate of absorption per unit area, analogous to **Error! Reference source not found.**,

$$\frac{dn_{2D}(y)}{dt} = \frac{2\pi}{\hbar} \frac{e^2}{4} E_{eff}^2(x_{QW}, y) r_{cv}^2 \rho_{2D}(\hbar\omega)[1 - f_c(E_c)] \tag{33}$$

Here the joint density of states per unit of energy per unit of area is $\rho_{2D}(\hbar\omega) = N_{QW} \frac{m_r}{\pi \hbar^2} H(\hbar\omega - E_g)$, where H is the step function, and the reduced mass $m_r$ is found from $m_r^{-1} = m_c^{-1} + m_v^{-1}$, and the value of energy in the conduction band (CB) is $E_c = (\hbar\omega - E_g) m_v / m_r$. Following (22) switching from the dipole to momentum representation we obtain

$$\frac{dn_{2D}(y)}{dt} = \frac{2\pi}{\hbar} \frac{e^2}{4} E_{eff}^2 \frac{1}{2} \frac{P_{cv}^2}{m_0^2 \omega^2} F_{cv} N_{QW} \frac{m_r}{\pi \hbar^2} H(\hbar\omega - E_g)[1 - f_c] \tag{34}$$

Next, in QWs the effective mass of the electron is determined by (23) while the in-plane effective mass of hole is typically about 2-3 times larger, then one can write for the reduced mass $m_r^{-1} = (1 + \beta_v) \times 2 P_{cv}^2 / m_0^2 \hbar\omega$ where factor $0 < \beta_v < 0.5$. Upon substituting it into (34) we obtain

$$\frac{dn_{2D}(y)}{dt} = \frac{1}{1 + \beta_v} \frac{e^2}{4\hbar} \frac{E_{eff}^2}{2\hbar\omega} F_{cv} N_{QW} H(\hbar\omega - E_g)[1 - f_c] \tag{35}$$

Next, following (1)-(5) we obtain the expression for the absorption coefficient

$$\alpha_{QW} = \frac{1}{1 + \beta_v} \frac{\pi \alpha_0}{n_{eff}} \frac{WFN_{QW}}{S_{eff}} F_{cv} H(\hbar\omega - E_g)[1 - f_c] \tag{36}$$

where $F$ is still described by (8) and the effective thickness is $t_{eff} = S_{eff}/WF$, same as (10). Next we want to evaluate the change in absorption occurring when the 2D density of carriers changes. Differentiating (20) over the electron density

$$\frac{d\alpha_{QW}}{dn_{2D}} = \frac{\partial \alpha_{QW}}{\partial f_c} \frac{\partial f_c}{\partial E_{fc}} \frac{dE_f}{dn_{2D}} = -\sigma_{QW}(\omega) t_{eff}^{-1} \tag{37}$$

where the differential absorption cross-section is

$$\sigma_{QW}(\omega) = \frac{F_{cv}}{1+\beta_v} \frac{\pi\alpha_0}{n_{eff}} \frac{\pi h^2}{m_c kT} \frac{\exp\left[(E_c - E_f)/kT\right]\left[1+\exp(-E_f/kT)\right]}{\left\{1+\exp\left[(E_c - E_f)/kT\right]\right\}^2} \quad (38)$$

and it reaches its maximum value when both the photon energy is equal to the bandgap and the Fermi level is also at the band-edge, i.e. $E_c = E_{fc} = 0$, resulting in:

$$\sigma_{QW}(\omega) = \frac{F_{cv}}{1+\beta_v} \frac{\pi\alpha_0}{n_{eff}} \frac{\pi h^2}{2m_c kT} \approx 2.75 \times 10^{-17} cm^2 \frac{1}{kT n_{eff}} \frac{m_0}{m_c} \frac{F_{cv}}{1+\beta_v} \quad (39)$$

The result is nearly identical to (25) – the only difference is that $kT$ plays the role of broadening. Note that the result does not depend on the number of QWs. Then we can proceed to estimate the change of absorption as charge can be related to the absorption as $\Delta\alpha_{QW} = \sigma_{QW}(\omega)\delta n_{2D}/t_{eff}$. This, however, will be an overestimation of the effect because we have linearized the dependence of absorption on density near $E_f = 0$ in (37). Approaching the problem from a different angle by following discussion in section 3.1. The minimum transmission (maximum absorption) is achieved when no electrons are injected:

$$\alpha_{QW,\max}(\omega)L = \frac{F_{cv}}{1+\beta_v} \frac{\pi\alpha_0}{n_{eff}} \frac{N_{QW}L}{t_{eff}}[1-f_{c,\min}] = \ln(10) \approx 2.302 \quad (40)$$

And, similarly to the 2-level case before, the maximum transmission (minimum absorption) is achieved when $\delta n_{2D}$ carriers are induced by the gate (injected).

$$\alpha_{QW,\min}(\omega)L = \frac{F_{cv}}{1+\beta_v} \frac{\pi\alpha_0}{n_{eff}} \frac{N_{QW}L}{t_{eff}}[1-f_{c,\max}] = -\ln(0.9) \approx 0.105 \quad (41)$$

Now, let us assume that the material is intrinsic and $E_{f,\min} \approx -E_{gap}/2$ therefore $f_{c,\min}(E_c = 0) = [1+\exp(-E_{f,\min}/kT)]^{-1} \approx 0$. The exact value of $f_{c,\min}$ does not influence the modulation as long as it is small. Then we obtain from (40), $\frac{F_{cv}}{1+\beta_v} \frac{\pi\alpha_0}{n_{eff}} \frac{N_{QW}L}{t_{eff}} \approx 2.3$. Substituting this into (41) we obtain $f_{c,\max}(E_c = 0) = [1+\exp(-E_{f,\max}/kT)]^{-1} \approx 0.96$ and $E_{f,\max} \approx 3.15kT$. According to (31), we can find $n_{2D,\max} \approx 3.2 N_{QW} m_c kT/\pi h^2$ while $n_{2D,\min} \approx 0$. Therefore, switching charge can be found as

$$Q_{SW} = eWL(n_{2D,\max} - n_{2D,\min}) = eW \times 2.2 \frac{t_{eff}}{N_Q} \frac{n_{eff}}{\pi\alpha_0} \frac{1+\beta_v}{F_{cv}} \times 3.2 \frac{m_c kT}{\pi h^2} N_{QW} \approx 2.2e \frac{S_{eff}}{\sigma_{QW}(\omega)F} \quad (42)$$

which is exactly the same as (15), as long as we can formally re-introduce effective differential cross-section (39) as

$$\sigma_{QW}(\omega) = \frac{F_{cv}}{1+\beta_v} \frac{\pi\alpha_0}{n_{eff}} \frac{\pi h^2}{m_c 3kT} \approx 5.4 \times 10^{-17} cm^2 \frac{1}{3kT n_{eff}} \frac{m_0}{m_c} \frac{F_{cv}}{1+\beta_v} \quad (43)$$

Finally one can combine the thermal broadening $3kT$ and the broadening $\gamma$ to an effective broadening $h\gamma_{eff} = \sqrt{(h\gamma)^2 + (3kT)^2}$ and modify (43) as

$$\sigma_{QW}(\omega) = \frac{1}{1+\beta_v} \frac{\pi^2\alpha_0}{n_{eff}} \frac{h^2}{m_c h\gamma_{eff}} F_{cv} \approx 5.4 \times 10^{-17} cm^2 \frac{1}{h\gamma_{eff} n_{eff}} \frac{m_0}{m_c} \frac{1}{1+\beta_v} F_{cv} \quad (44)$$

which has the same form as (24) producing similar results. However, when choosing the modulator length we now have less flexibility to assure that the maximum (off-state) absorption is satisfied according to

$$L_{QW} \approx 2.3 \frac{n_{eff} t_{eff}}{\pi \alpha_0 N_{QW}} \frac{1+\beta_v}{F_{cv}} \tag{45}$$

The only adjustable parameter is the number of QWs, which cannot be too large because the carriers will always tend to accumulate non-uniformly in the few QWs closest to the gate. In this respect, the Pauli-blocking type of modulator is inferior to the QCSE modulators in which no charge is stored. If we compare this with the results for QDs (26) we can see that QD modulator would have the same length if the density of QDs is $N_{QD} \approx \gamma m_c / \hbar \approx 2 \times 10^{12} cm^{-2}$ which is a bit too high for the self-organized QDs – hence QW modulator has advantage of shorter length, which increases the speed yet increases switching voltage.

## 3.4 Graphene

Even though there have been many derivations of graphene absorption, it is instructive to re-derive the inter-band absorption in graphene from the first principle because it allows us to see how closely related it is to absorption by any other 2D structure whether there are any Dirac electrons involved in it. In fact, when the electrons involved in Pauli blocking are located at roughly 0.4eV away from the Dirak point, all the peculiarity of graphene becomes irrelevant and as a result its performance as an electro absorption modulator is no different from any other semiconductor material. In graphene, the matrix element of momentum between valence and conduction band is $P_{cv,gr} = m_0 v_F$, where $v_F \sim 10^8 cms^{-1}$ is the Fermi velocity. The matrix element of Hamiltonian in the $\mathbf{p} \cdot \mathbf{A}$ gauge then becomes $H_{cv} = \frac{1}{2} e \mathbf{v}_F \cdot \mathbf{E} / \omega$. The Joint density of states can be found next, as

$$\rho_{2D}(\hbar \omega) = \frac{1}{2\pi} \frac{\omega}{\hbar v_F^2} \tag{46}$$

where we have used the dispersion relation for the transition frequency, $\omega = 2k v_F$ and the fact that in graphene one deals with both spin and valley (K,K') degeneracies. Substituting it all into the Fermi golden rule we obtain:

$$\frac{dn_{2D}(y)}{dt} = \frac{2\pi}{\hbar} \frac{e^2}{4} E_{eff}^2 \frac{1}{2} \frac{v_F^2}{\omega^2} \frac{1}{2\pi} \frac{\omega}{\hbar v_F^2}[1-f_c] = \frac{e^2 E_{eff}^2}{8\hbar^2 \omega}[1-f_c] \tag{47}$$

This expression is quite similar to (35) and thus following all the steps for QWs that have lead to (36) we obtain the expression for the graphene interband absorption in the waveguide:

$$\alpha_{gr} = \frac{\pi \alpha_0}{n_{eff} t_{eff}}[1-f_c] \tag{48}$$

Next, we consider the changes in graphene absorption with density of electrons. Since the changes occur when the Fermi level approaches $E_f = \hbar \omega / 2 \gg kT$, the relation between density of electrons and Fermi level is $n_{2D} \approx E_f^2 / \pi \hbar^2 v_F^2$ and differentiating it we obtain $dn_{2D} / dE_f \approx 2E_f / \pi \hbar^2 v_F^2$. Now, following (37)

$$\frac{d\alpha_{gr}}{dn_{2D}} = \frac{\partial \alpha_{QW}}{\partial f_c} \frac{\partial f_c}{\partial E_{fc}} \frac{dE_f}{dn_{2D}} = -\sigma_{gr}(\omega) t_{eff}^{-1} \tag{49}$$

Where the differential absorption cross-section is

$$\sigma_{gr}(\omega) = \frac{\pi^2 \alpha_0}{n_{eff}} \times \frac{1}{kT} \frac{\exp\left[(h\omega/2 - E_{fc})/kT\right]}{\left\{1 + \exp\left[(h\omega/2 - E_{fc})/kT\right]\right\}^2} \frac{h^2 v_F^2}{2E_f} \tag{50}$$

This expression obviously peaks at $E_f = h\omega/2$ and we obtain

$$\sigma'_{gr}(\omega) = \frac{\pi^2 \alpha_0}{n_{eff}} \times \frac{1}{kT} \frac{1}{4} \frac{h^2 v_F^2}{h\omega} \approx 9.7 \times 10^{-17} cm^2 \frac{1}{n_{eff} kT} \tag{51}$$

Finally, similar to the previous section, in order to account for the fact that the absorption does not change linearly with the Fermi level and also for the additional broadening the effective broadening is introduced as b $h\gamma_{eff} = \sqrt{(h\gamma)^2 + (3kT)^2}$, and

$$\sigma'_{gr}(\omega) = \frac{\pi^2 \alpha_0}{n_a} \times \frac{1}{h\gamma_{eff}} \frac{1}{4} \frac{h^2 v_F^2}{h\omega} \approx 9.7 \times 10^{-17} cm^2 \frac{1}{h\gamma_{eff}} \tag{52}$$

Comparing (52) with (44) we see that for a bandgap of ~0.8 eV the effective mass is about $m_c \approx 0.05 m_0$. Assuming $\beta_v = 1$, the results are strikingly similar for graphene and QWs which is easy to understand since graphene and many typical III-V semiconductors have roughly similar matrix elements of transition, and for a given transition energy have similar joint densities of states.

### *3.5 Excitons in 2D Semiconductors*

We now turn our attention to another prospective medium for electro-absorption modulators which has recently seen renewed interest – two dimensional (2D) excitons. Excitons in the 2D semiconductor QWs was considered back in 1980's, mostly in the context of nonlinear optical devices [8]. But the binding energies in the III-V semiconductor QWs are comparable to room temperature thermal energy, hence excitons easily dissociate and have a broad absorption spectrum, which in the end made QW devices based on excitonic absorption impractical. Yet in the last decade the interest has shifted to the 2D TMDs where the low dimensionality of the material increased the binding energy, mostly due to reduced Coulomb-Coulomb screening, as compared to III-Vs. Thus, TMDs show increased absorption over QWs, and it was further suggested that these new 'robust' excitons can be used for efficient light modulation [9]. However a higher 'robustness' of the exciton should make it more difficult to change its absorption by any means, be that states saturation or screening. The 2D material exciton is characterized by its 2D Bohr radius

$$a_{ex} = \frac{2\pi \varepsilon_{eff} \varepsilon_0 h^2}{e^2 m_r} \tag{53}$$

which is half as large as 3D exciton radius. The effective dielectric constant $\varepsilon_{eff}$ in 2D materials approaches unity while the effective mass $m_r \sim 0.25 m_0$ is somewhat larger than in III-V semiconductors. Thus, the exciton Bohr radius in TMDs is on the order of only a few nm [10], exhibiting exciton binding energy of $E_{ex} = h^2 / 2m_r a_{ex}^2 \sim 0.5 eV$ [11]. The absorption of the exciton can be easily obtained by using the value of the exciton envelope wave function at the origin, i.e. probability of finding electron and hole in the same spatial location $2|\Phi_{ex}(0)|^2 = 4/\pi a_{ex}^2$ in place of two 2D density of 'atoms' or 'QDs' in (9) to obtain

$$\alpha_{ex}(\omega) = 4\sigma_{ex}(\omega)/\pi a_{ex}^2 t'_{eff} \qquad (41)$$

where according to (4) and (24) operating on the excitonic resonance

$$\sigma_{ex}(\omega) = 4\pi\alpha_0 r_{12}^2 \frac{\omega}{\gamma} \approx 2\pi\alpha_0 \frac{\text{h}}{\gamma}(m_c^{-1} - m_0^{-1}) \qquad (42)$$

where we have used $r_{12}^2 = P_{cv}^2 / m_0^2 \omega^2 = \text{h}/2m_c\omega$. The absorption is quite large (Fig. 2) and therefore the length of the modulator, according to (8) can be quite small, $L_{ex} \approx 2.302 t_{eff} a_{ex}^2 / 4\pi\sigma_{ex}(\omega) : 10 t_{eff}$. Next, we inject free carriers into the TMDs; where three processes take place simultaneously: (i) state filling, (ii) bandgap renormalization and (iii) screening. Thus, excitons bleach as recently observed by Heinz who attributed it Mott transition [10]. Mott transitions occur in 3D semiconductor when the screening radius becomes comparable to the exciton radius, i.e. when $n_{2D} \sim a_{ex}^{-2}$. Strictly speaking the screening in 2D system saturates. Therefore, exciton bleaching most likely takes place because of state filling [8]. The exciton wave function can be considered a coherent superposition of the electron-hole-pair states with the wave vectors between 0 and roughly $1/a_{ex}$. The density of these states (with spin and valley degeneracy) is roughly $n_{bl} : 4/\pi a_{ex}^2$. The exact dependence of excitonic absorption on the density of injected carriers may be quite difficult, however, for our order-of-magnitude analysis it can be linearized as

$$\alpha(\omega, n_{2D}) = \alpha(\omega, 0) \times (1 - n_{2D}/n_{bl}) = \frac{8\alpha_0}{a_{ex}^2 t'_{eff}} \frac{\text{h}}{m_c \gamma}\left(1 - \pi a_{ex}^2 n_{2D}/4\right) = \frac{8\alpha_0}{a_{ex}^2 t_{eff}} \frac{\text{h}}{m_c \gamma} - \sigma'_{ex} n_{2D} \quad (54)$$

As one can see, the exciton radius and binding energy do not play a role in the determination of the switching charge, which is still determined by the expression (15) with the effective cross section for the exciton as

$$\sigma'_{ex}(\omega) = 2\pi\alpha_0 \frac{\text{h}}{m_c \gamma_{ex}} \approx 3.53 \times 10^{-17} cm^2 \frac{1}{\text{h}\gamma_{ex}}\left(\frac{m_0}{m_c} - 1\right) \qquad (55)$$

which is comparable to the QDs. Since the effective mass in TMDs is typically larger than in III-V semiconductors, it appears that using excitons does not change the fundamental fact that each time a single electron is injected inside the active layer a single transition is being blocked. Given the fact that the oscillator strength for each allowed transition is roughly the same, no matter what material is used, the change of absorption is expected to be constant.

### 4. Material Absorption Characteristics

All the absorption cross sections for different material classes are summarized in Table 1. The main observation that can be made here is that for all the materials operating with Pauli blocking the results are roughly comparable, and the free carriers offer far worse performance due to non-resonant character of absorption since according to (33) for free carriers $\gamma_{eff} = \gamma + \omega^2/\gamma > 2\omega$. To verify these analytical results, obtained from essentially perturbative approach, we calculate the dependence of the absorption on the injected (induced by the gate) carriers $n_{2D}$ for all the materials. For the QWs we obtain according to (36) and (31),

**Table 1. Summary of the absorption cross sections for different material classes**

| Material Absorption Cross-section | Expression | Approximate result |
|---|---|---|
| QD | $\sigma_{QD}(\omega) = \dfrac{\pi\alpha_0}{n_{eff}} F_{cv} \dfrac{h^2/m_0}{h\gamma}\left(\dfrac{m_0}{m_c}-1\right)$ | $1.8\times 10^{-17} cm^2 \dfrac{F_{cv}}{h\gamma n_{eff}}\left(\dfrac{m_0}{m_c}-1\right)$ |
| QW | $\sigma'_{QW}(\omega) = \dfrac{F_{cv}}{1+\beta_v}\pi^2\alpha_0 \dfrac{h^2}{m_c h\gamma_{eff}}$ | $5.4\times 10^{-17} cm^2 \dfrac{1}{h\gamma_{eff}}\dfrac{m_0}{m_c}\dfrac{F_{cv}}{1+\beta_v}$ |
| Graphene | $\sigma'_{gr}(\omega) = \pi^2\alpha_0 \times \dfrac{1}{h\gamma_{eff}}\dfrac{1}{4}\dfrac{h^2 v_F^2}{h\omega}$ | $9.7\times 10^{-17} cm^2 \dfrac{1}{h\gamma_{eff}}$ |
| WSe$_2$ | $\sigma'_{ex}(\omega) = 2\pi\alpha_0 \dfrac{h}{m_c \gamma_{ex}}$ | $3.5\times 10^{-17} cm^2 \dfrac{1}{h\gamma_{ex}}\dfrac{m_0}{m_c}$ |
| Free Carriers (at optimum $\gamma$) | $\sigma'_{fc,max}(\omega) = 4\pi\alpha_0 \dfrac{h}{m_c \gamma_{eff}}$ | $7.1\times 10^{-17} cm^2 \dfrac{1}{h\gamma_{eff}}\dfrac{m_0}{m_c}$ |

$$\alpha_{QW} = \frac{\pi\alpha_0}{n_{eff} t_{eff}}\frac{F_{cv}}{1+\beta_v} N_{QW}\left(\left[\exp\left(\frac{\pi h^2 n_{2D}}{N_{QW} m_c kT}\right)-1\right]e^{-\frac{E_c}{kT}}+1\right)^{-1} \quad (56)$$

which is plotted in Fig. 2(a) in units of $1/n_{eff}t_{eff}$ for $N_{QW} = 1$ and 3, $m_c = 0.06 m_0$, $E_c \approx h\omega - E_g = 50 meV$, $F_{cv} = 0.8$, $kT = 300K$, $\beta_v = 0.4$. Similarly for graphene, following **Error! Reference source not found.** and **Error! Reference source not found.**

$$\alpha_{gr} = \frac{\pi\alpha_0}{n_{eff} t_{eff}} \frac{1}{e^{\frac{hv_F\sqrt{\pi n_{2D}}-h\omega/2}{kT}}+1} \quad (57)$$

For QDs according to (24) we estimate

$$\alpha_{QD} = \frac{\pi\alpha_0}{n_{eff} t_{eff}} F_{cv} \frac{h^2 N_{QD}}{m_0 kT}\frac{kT}{h\gamma_{eff}}\left(\frac{m_0}{m_c}-1\right)\left(1-\frac{n_{2D}}{N_{QD}}\right) \quad (58)$$

The results are shown in Fig. 2(a) for the same effective mass and overlap as QWs, $h\gamma = kT$, $N_{QD} = 10^{12} cm^{-2}$. For the excitons, according to **Error! Reference source not found.**

$$\alpha_{ex}(\omega) = \frac{8\alpha_0}{n_{eff} t_{eff}}\frac{h^2 a_{ex}^{-2}}{m_0 kT}\frac{kT}{h\gamma_{eff}}\frac{m_0}{m_c}\left(1-\pi a_{ex}^2 n_{2D}/4\right) \quad (59)$$

The results are shown in Fig. 2(a) using exciton radius, $a_{ex} = 2nm$ and the conduction effective mass, $m_c = 0.2 m_0$. For the free carrier modulators, we use

$$\alpha_{fc}(\omega) = \frac{4\pi\alpha_0}{n_{eff}t_{eff}} \frac{\hbar}{m_c} \frac{\gamma}{\omega^2 + \gamma^2} n_{2D} \tag{60}$$

$\gamma = 1/\tau$ is the carrier scattering rate i.e. collision frequency, electron mobility $\mu$ and $\tau$ are related by $\mu = |q|\tau/m_c$. The conductivity effective mass, $m_c$ is taken as $0.26m_0$ [12]. $\mu$ is taken as 1100 cm$^2$V$^{-1}$s$^{-1}$ at $10^{16}$ cm$^{-3}$ carrier concentration level (i.e. electrons for Silicon) [13]. Unlike the doped silicon, ITO whose chemical composition is usually given as $In_2O_3:SnO_2$ can be considered an alloy as concentration of tin relative to indium can be as high as 10%. Several previous studies have calculated the permittivity of ITO using the experimentally measured reflectance and transmittance, and we choose a fitting result of Michelotti et. al whereas $\gamma$ depends on the deposition conditions, in our analysis we have taken $\gamma = 1.8 \times 10^{14}$ rad s$^{-1}$ [14-17].

Looking at the results in Fig. 2(a), where the normalized absorption $\alpha' = \alpha n_{eff} t_{eff}$ is plotted, one cannot help but notice that while the absolute values of absorption are different for all Pauli-blocking schemes, the all-important slope of the curves are very similar, as is expected from the similarity of the differential absorption cross-sections in Table 1. Next, we find the length required to achieve ~10dB off state absorption in units of $t_{eff}$ i.e. $L' = L/t_{eff}$ according to (16) (Fig. 2d). Furthermore, we evaluate the change of total absorption $\alpha L = \alpha' L'/n_{eff}$ as a function of the injected charge per unit waveguide cross-section $Q' = Q/S'_{eff} = en_{2D}L't_{eff}W/S'_{eff} = en_{2D}L'$ (Fig. 2b). For graphene, we only show the AC charge $Q' = e(n_{2D} - n_{2D,0})L'$ where $n_{2D,0} = 0.9 \times 10^{13}$ cm$^{-2}$ is the electron density that brings the Fermi level within $3kT$ from the 0.4eV. From Fig. 2(b), we can determine the switching charge necessary to obtain 10dB on off ratio – the values of switching charges corresponding to the material classes for 10 db modulation is shown in Fig. 2(e). According to our theoretical estimate, they all (with the exception of free carrier schemes) should lie within the range of $Q' \sim 2.2e/\sigma \sim 10^{-13} - 10^{-12} C/\mu m^2$ and this appears to be true. The difference between graphene, QWs, QDs and TMDs is not significant considering that the exact amount of broadening cannot be speculated. So, in the end, it all depends on the effective cross-section, and the only distinction between the materials is how easily they can be integrated into the small waveguide.

Next, we calculate the capacitance per $\mu m^2$ of the effective cross section $C'_g = C_g/S'_{eff} = \varepsilon_0 \varepsilon_{eff} L'/d_{gate}$ (Fig. 2f). The spacing between the active layer and the gate is $d_{gate} = 100nm$ and $\varepsilon_{eff} = 10$ assumed in our calculations. This allows us to obtain the drive voltage via $V_d = Q/C_g = Q'/C'_g$ so that the absorption vs. drive voltage can be plotted in Fig. 2(c); and from there switching voltage can be obtained (Fig. 2j). Note, switching voltage does not depend on the waveguide geometry, only on the gate insulator thickness. In terms of switching voltage, the QD modulator appears superior, but at the expense of much larger capacitance caused by the fact that density of QDs is quite low and long length is required to achieve 10dB attenuation. Now, one can finally determine the switching energy per bit per unit effective waveguide area, as $U'_{sw} = U_{sw}/S'_{eff} = \frac{1}{2}Q'_{sw}V_{sw} = \frac{1}{2}Q'^2_{sw}/C'_g$ (Fig. 2g).

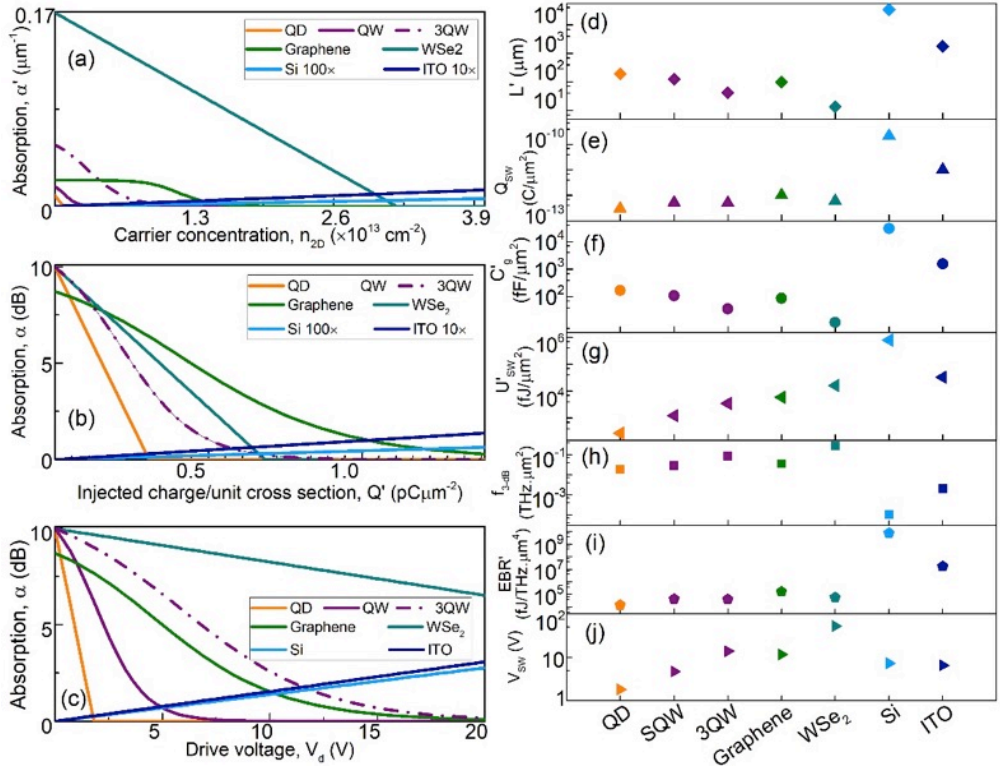

**Fig. 2:** Absorption modulation results for different material classes, for charge-driven active materials. **(a)** Normalized absorption, $\alpha' = \alpha n_{eff} t_{eff}$ as a function of carrier concentration, $n_{2D}$ (cm$^{-2}$); **(b-c)** Optical absorption vs. **(b)** injected charge, $Q'$ (pC$\mu$m$^{-2}$); and **(c)** drive voltage, $V_d$ (Volts); respectively. **(d-j)** Physical device performance parameters to obtain 10dB modulation; **(d)** Modulator length, $L'=L/t_{eff}$; **(e)** Switching charge, $Q_{SW}$ (C/$\mu$m$^2$); **(f)** Electrical device capacitance, $C'_g$ (fF/$\mu$m$^2$); **(g)** Switching energy (energy-per-bit function), $U'_{SW}$ (fJ/$\mu$m$^2$); **(h)** 3-dB modulation speed, $f_{3dB}$ (THz.$\mu$m$^2$); **(i)** Energy-bandwidth ratio, $EBR'$ (fJ/Thz.$\mu$m$^4$); and **(j)** Switching voltage, $V_{SW}$ (Volts) for investigated material classes including quantum dots (QD), single and 3-layer quantum well (SQW, 3QW), Graphene, a transition metal dichalcogenide (TMD) material (WSe$_2$), Silicon, and Indium-Tin-Oxide (ITO).

Also, the 3dB cut off frequency $f_{3dB} = 1/2\pi R C_g$ can be calculated under assumption of $R = 50\Omega$ (Fig. 2h). This is where single layer TMDs appear competent, because large excitonic absorption allows one to use a very short path length, hence low capacitance. But of course, as is obvious from Fig. 2(c), the driving voltage for such modulator is huge as it is very difficult to saturate this strong absorption. We note, however, that modal implications, in particular, in photonic waveguides, typically result in higher contact resistances than $50\Omega$ [18]. Plasmonics-based devices allow defining the electrical capacitor with high spatial overlap to the actual device region [19].

The relevant figure of merit (FOM), for such modulators, is the ratio of the switching energy and cut-off frequency ("Energy Bandwidth Ratio" or *EBR*),

$$EBR = U_{SW} / f_{3dB} = \pi Q_{SW}^2 R \qquad (67)$$

Where evidently lower *EBR* is desired. In Fig. 2(i), we plot $EBR' = EBR / S_{eff}^{'2} = \pi Q_{SW}^{'2} R$ in units of $fJ/(THz \cdot \mu m^4)$. It can be noticed that for all the materials that do not rely on off-resonant absorption of free carriers the *EBR'* does not stray far from the same value of roughly

$10^5 \, fJ/(THz \cdot \mu m^4)$, as is expected since the switching charges (Fig. 2e) are in very close range. Using the switching charge in (19), we find that

$$EBR = \pi Q_{SW}^2 R \sim \pi e^2 RS_{eff}^{'2} \times \left(\frac{m_c}{m_0}\right)^2 \left(\hbar\gamma_{eff}\right)^2 \times 2\times 10^{17} \, \mu m^{-4}$$

$$\approx 5\times 10^4 \, \frac{S_{eff}^{'2}}{\mu m^4} \times \left(\frac{m_c}{0.067 m_0}\right)^2 \left(\frac{\hbar\gamma_{eff}}{100 meV}\right)^2 \, fJ/THz \quad (68)$$

for approximately all the materials that do not rely on free carrier absorption. Note, within the constraints of (68), one can increase the bandwidth by increasing the thickness or dielectric constant of the insulator while also increasing the switching voltage and energy – in the same way how one would adjust the threshold voltage and speed of the field effect transistor (FET). Of course, it is important to include parasitic capacitances and (68) simply presents the fundamental upper bound of the FOM. Since the effective mass in the conduction band is determined by the bandgap (larger the bandgap, the larger the effective mass roughly), the only "intrinsic" material property affecting device performance is the effective broadening, which for most systems cannot be reduced below few *kT* without external cooling. This only leaves the external parameter – effective cross section of the waveguide as the main factor for consideration which is capable of improving the FOM. We note that the intrinsic disadvantages of 2D materials being atomically thin may be compensated by achieving smaller cross-sections using, for instance, plasmonic modes.

## 5. Effective Area for Different Modal Structures

Here we choose 14 different mode structures with different materials and photonic/plasmonic designs (Fig. 3a-3n). Representative of the free carriers, we choose conventional Silicon (Si) and a transparent conductive oxide (TCO) emerging material, Indium Tin Oxide (ITO). We have included Graphene as a unique 2D material, WSe$_2$ representing TMDs, and InGaAs representing conventional III-V materials. In order to understand the impact on the effective area, we diversify photonic and plasmonic waveguide mode designs to include different mode structures, such as bulk, slot [20] and hybrid-photon-plasmon (HPP) [21,22] for the chosen materials – Si, ITO, Graphene and WSe$_2$. The Si, ITO and Graphene mode structures are chosen similar to our previous work [23,24]. The WSe$_2$ structures are the same as graphene ones with just the active material changed to WSe$_2$. In principle, III-V QWs or QDs can also be placed inside the slot structure also, but such designs are not considered in this work. As expected, bulk waveguide designs do not allow for small effective cross-sections. However, the actual improvement is not more than 100× expect for the Silicon slot case where the mode is highly squeezed due to the high index Si being below the slot air gap realizing the high E-field concentration, but such a miniscule effective area in this mode also presents tradeoff in terms of high insertion loss of 27 dB. We note that the waveguide dimensions assumed in Figure 3 are exemplary, but smaller effective cross-section are obtainable as well. Yet, a balance between insertion loss and shrinking the mode size should be considered from a link integration point of view [25]. Employing plasmonic structures in this regard can help the cause as plasmonic modes can squeeze in the light by a few orders of magnitude [26-33].

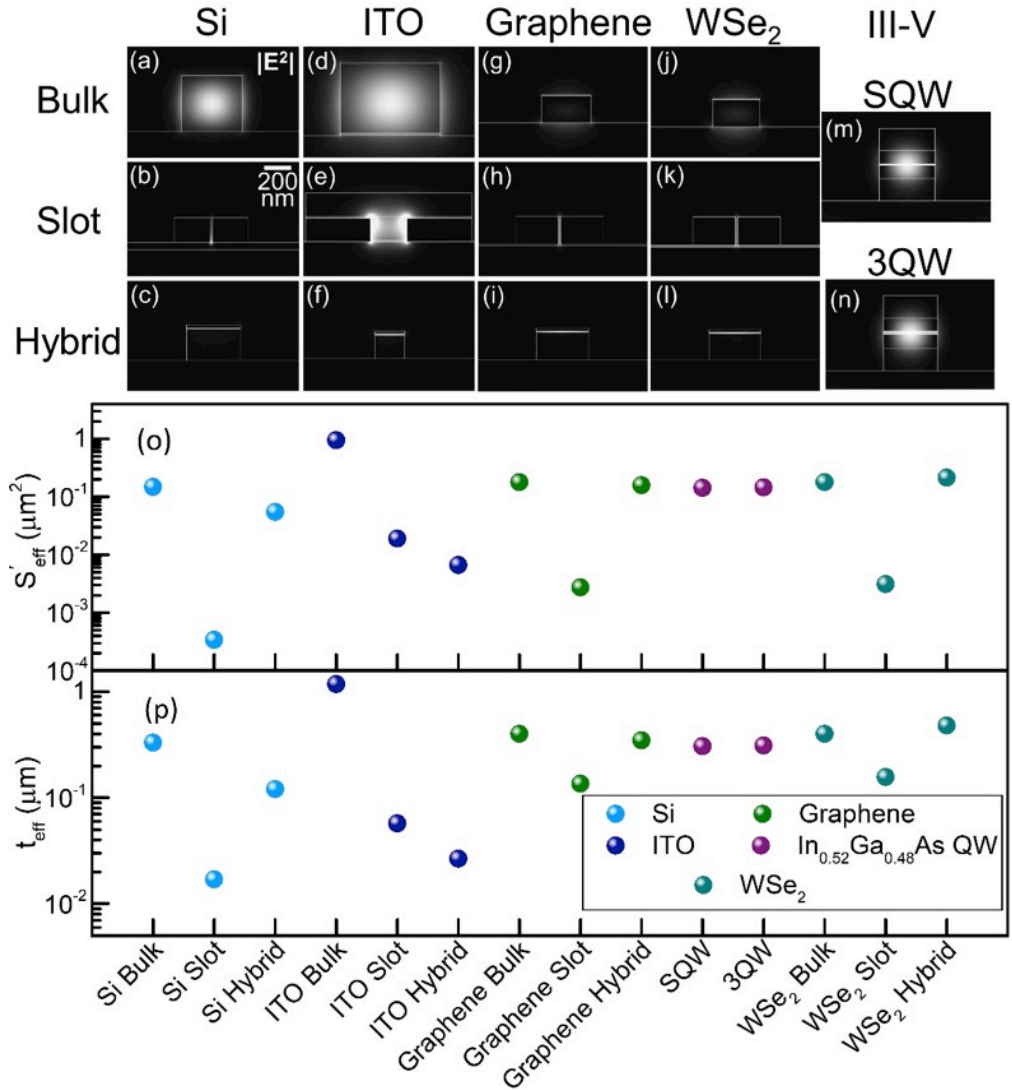

**Fig. 3: (a-n)** Cross-sectional mode profiles for different structures and material classes using FEM analyses, normalized electric field intensity, $|\mathbf{E}|^2$ is shown. **(o)** Corresponding effective cross-sectional modal area, $S'_{eff}$; and **(p)** effective thickness, $t_{eff}$. The Si, ITO and graphene structures are chosen form our previous work [23,24]. WSe$_2$ structures are the same in dimensions as graphene ones just changing the active material to WSe$_2$, In$_{0.52}$Ga$_{0.48}$As QWs are chosen having 5 nm thickness and 470 nm width, with GaAs separate confinement heterostructure (SCH) and barrier layers.

## 6. Energy-per-bit Function, Speed and Material Broadening

As we have mentioned the free carrier modulators operate far from resonance and are therefore not competitive with the other material structures. In fact, reducing the broadening will only make their performance as an electro-absorption modulator even worse (which is not the case for the electro-optic modulators). But the other transition blocking based material systems do use resonant absorption, hence can benefit from reduced broadening as the transitions are sharper and a higher ER is possible in the same footprint. In case of QDs and TMDs the absorption linewidth contains homogeneous part which is at least partially temperature-dependent (phonon scattering) and the inhomogeneous one due to material variations (especially in QDs). In the graphene and QWs, in addition to scattering, the main cause of broadening is associated with the Fermi-function spread on the scale of *3kT*. For these materials the effective linewidth is $\gamma_{eff} = \sqrt{\gamma_{homo}^2 + \min\left\{\begin{array}{c}3kT \\ \text{or} \\ \gamma_{inh}\end{array}\right\}^2}$ , where $\gamma_{homo}$ is the material or homogeneous broadening and $\gamma_{inh}$ is the inhomogeneous broadening. All the EAM performance parameters with respect to the effective broadening, effective thickness and effective cross-sectional area are calculated (Fig. 4).

Following previous discussions, we include different material classes to demonstrate the effect of effective material broadening and effective modal area variation. Pauli blocking based graphene, band filling based QW modulators, narrow resonant QDs, and excitonic modulation based WSe2 are chosen to investigate effects of changing the effective broadening. Different free carrier based materials, namely Si and ITO, are also chosen to signify variation of effective material broadening in free carrier based schemes. Our results show monotonic modulator improvements with both reduced broadening and effective area. The latter points to polaritonic modes where the field is squeezed into small (i.e. sub-diffraction limited) effective modal areas. Regarding broadening, almost 1-2 orders of magnitude improvements can be attained by operating at cryogenic temperatures, i.e. $\gamma \ll$ 0.1eV. We used a nominal 77 K for our calculations corresponding to cryogenic temperature. However, beyond a broadening corresponding to room temperature, the effect declines. To study the dependence on the effective broadening, $\gamma_{eff}$ for QWs, and graphene while we change the temperature the device parameters change with the scaling factor, $G = T/300\ K$. As such, the switching charge, $Q_{SW}$ scales as $G$; capacitance, $C$ stays the same; as a result the 3-dB modulation bandwidth stays the same also. The modulator length, $L$ is unchanged from the variation in broadening; the switching voltage, $V_{SW}$ scales as $G$; switching energy, i.e. the energy-per-bit function, $U_{SW}$ and energy-bandwidth ratio, $EBR$ both scale as $G^2$. The situation is different for resonant narrow line emitters like QDs and WSe2. Since we used $\gamma = kT$ for them in previous discussions, we introduce the scaling factor, $G = \gamma/kT$ in order to observe variations in changing $\gamma$. Incidentally, now broader $\gamma$ means longer length needed to absorb. Therefore, switching charge, $Q_{SW}$ scales as $G$; capacitance, $C$ scales as $G$; modulator length, $L$ scales as $G$; the switching voltage, $V_{SW}$ stays the same ($\sim Q_{SW}/C$); switching energy, $U_{SW}$ scales as $G$; and finally, energy-bandwidth ratio, $EBR$ scales as $G^2$.

As the 3-dB speed increases and the energy-per-bit decreases with the effective modal area, our deterministic figure of merit, *EBR* also decreases significantly by quite a few orders of magnitude; validating the claim for using compact and sub-diffraction limited modes (i.e. plasmonic modes, slot waveguides, QDs etc.) in order to achieve higher LMI and modulation performance.

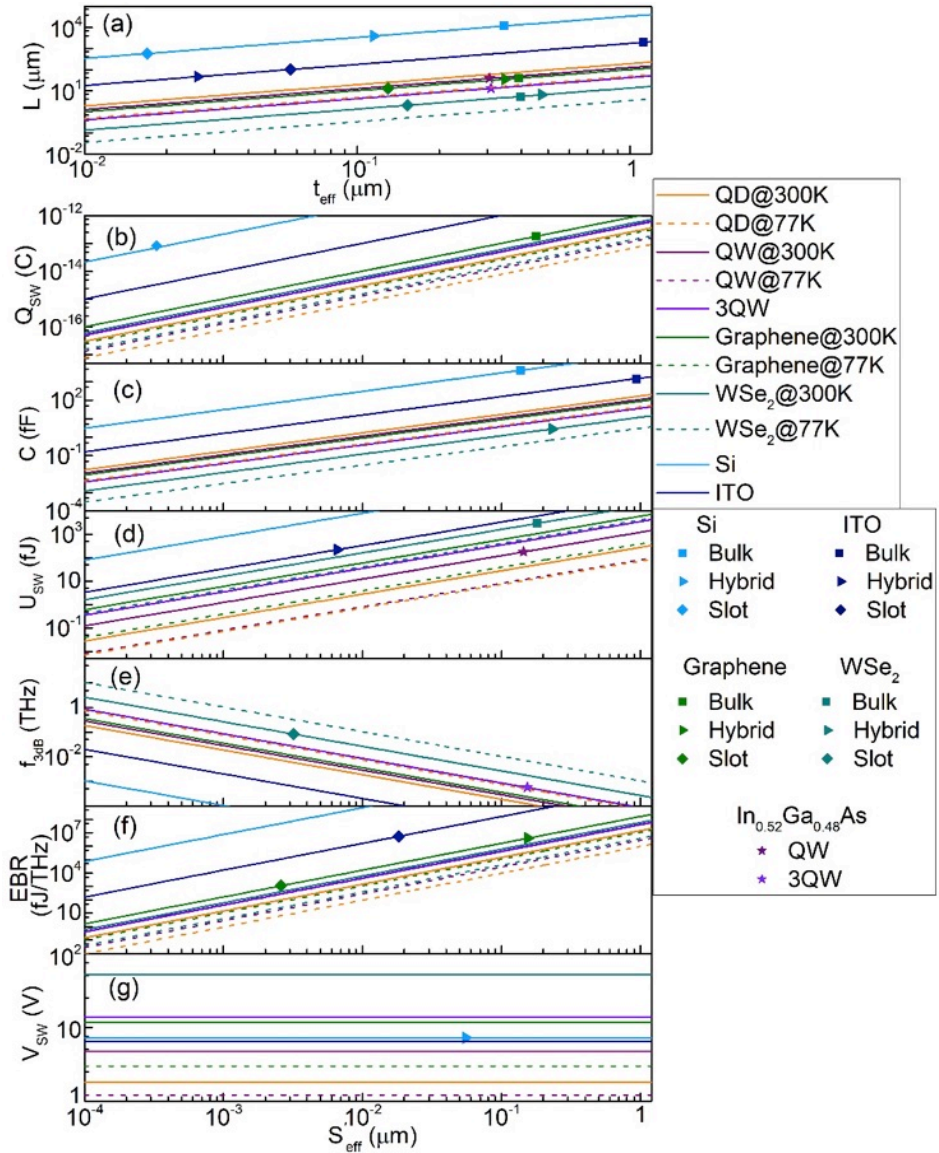

**Fig. 4: (a)** 10 dB absorption modulator length, *L* vs. effective thickness, $t_{eff}$ for all the comparable material classes with different broadening. **(b-g)** Relevant device parameters vs. effective mode areas, $S'_{eff}$ for different effective material broadening, $\gamma_{eff}$ (eV) for all the comparable material classes; **(b)** Switching charge, $Q_{SW}$ (C); **(c)** Capacitance, *C* (fF); **(d)** Switching energy (E/bit function), $U_{SW}$ (fJ); **(e)** 3-dB modulation bandwidth, $f_{3dB}$ (THz); **(f)** Energy-bandwidth ratio, *EBR* (fJ/THz); and **(g)** Switching voltage, $V_{SW}$ (V). The varying amount of broadening is implicit by the materials shown in the legend corresponding to room temperature, i.e. 300 K, and thermal cooling down to 77 K. The corresponding $t_{eff}$ and $S'_{eff}$ for the different modes from the previous section are also marked to aid comparing device performances. $t_{ox}$ = 100 nm. Lowering the oxide thickness will increase the energy efficiency through improved electrostatics, but also reduce the modulation speed via increased capacitance.

## 7. Conclusion

We carried out a holistic physical device analysis for waveguide-based electro-absorption modulator. Our analysis reveals an intricate dependency of obtainable modulator performance on the fundamental material class properties, and obtainable waveguide cross-section absorption. We include two-level absorbers such as quantum dots, free carrier accumulation or depletion such as ITO or Silicon, two-dimensional electron gas in semiconductors such as quantum wells, Pauli blocking in Graphene, and excitons in two-dimensional atomic layered materials found in transition metal dichalcogendies. Key results are that independent of the material class, the broadening effects of the material transition critically determines the amount of electrical charges needed for switching. Disregarding the free carriers all classes have essentially the same effective cross section, because for the allowed transitions the scale of the momentum matrix element scales inversely with the chemical bond length which does not change much from one material to another. Therefore, minimizing the broadened transition is key. However quantum dots always have the broadening in excess of 100 meV, and in graphene and quantum wells thermal broadening alone is about 75 meV at room temperature. Inversely, however, for free carriers broadening reduces the required switching voltage. Beyond broadening, what is left then is to consider are waveguide designs with smaller effective cross-section. Thermal cooling can be a valid approach to increase the modulation efficiencies as we progress into a more data driven world where the need for efficient modulation and detection in complex modulation schemes (e.g. 16/32 QAM) can probably outweigh relevant costs with thermal cooling.

## Acknowledgement

V.S. is supported by ARO (W911NF-16-2-0194), by AFOSR (FA9550-14-1-0215) and (FA9559-15-1-0447), which is part of the data-driven applications system (DDDAS) program.